\documentclass[aps,prb,twocolumn,showpacs,preprintnumbers,amsmath,amssymb]{revtex4}
\usepackage{graphicx}
\usepackage{dcolumn}
\usepackage{bm}
\usepackage{subfigure}
\usepackage{color}
\usepackage{amsmath}
\usepackage{amssymb}

\begin{document}


\title{Magnetic and Electrical Properties of DHCP NpPd$_3$ and (U$_{1-x}$Np$_x$)Pd$_3$}

\author{H. C. Walker}
 \author{K. A. McEwen}%
 \email{k.mcewen@ucl.ac.uk}
 \affiliation{Department of Physics and Astronomy, and London Centre for Nanotechnology, University College
 London, Gower Street, London, WC1E 6BT, UK}%

\author{P. Boulet}
\altaffiliation[present address: ]{Laboratoire de Sciences et
G\'{e}nie des Mat\'{e}riaux et de M\'{e}tallurgie, UMR 7584
CNRS-INPL, Ecole des Mines, Parc de Saurupt 54042 Nancy Cedex,
France}
\author{E. Colineau}
\author{J.-C. Griveau}
\author{J. Rebizant}
\author{F. Wastin}
 \affiliation{European Commission, Joint Research Centre, Institute for Transuranium Elements,
 Postfach 2340, Karlsruhe, D-76125 Germany}

\date{\today}

\begin{abstract}
We have made an extensive study of the magnetic and electrical
properties of double-hexagonal close-packed (DHCP) NpPd$_3$ and a
range of (U$_{1-x}$Np$_x$)Pd$_3$ compounds with
$x=0.01,\,0.02,\,0.05$ and $0.50$ using magnetisation, magnetic
susceptibility, electrical resistivity and heat capacity
measurements on polycrystalline samples, performed in the
temperature range $2-300$ K and in magnetic fields up to 9 T. Two
transitions are observed in NpPd$_3$ at $T=10$ K and 30 K. Dilute Np
samples ($x\leq0.05$) exhibit quadrupolar transitions, with the
transition temperatures reduced from those of pure UPd$_3$.
\end{abstract}

\maketitle

\section{Introduction}
The magnetism of $4f$ (lanthanide) metallic systems has been studied
extensively for several decades. The standard localised moment model
of rare earth magnetism, as expounded by Jensen and
Mackintosh\cite{Jensen}, remains the foundation for explaining the
magnetic and electrical properties of almost all such materials,
with the exception of cerium and its compounds, where band and
hybridisation effects play a crucial role. In $5f$ (actinide)
metallic systems, the $f$-electron wave functions are distinctly
more extended than in the corresponding $4f$ systems, leading to a
more itinerant-like nature as manifested by their magnetic and
electrical behaviour. Actinide systems display a unique complexity
within the periodic table, exhibiting a rich variety of phenomena
such as heavy-fermion and non-Fermi liquid behaviour, and
unconventional superconductivity.  Most uranium metallic compounds
have the U$^{3+}$ $5f^3$ configuration and exhibit itinerant-like
properties. In contrast, UPd$_3$ shows the U$^{4+}$ $5f^2$
configuration, and is a rare example of a localised moment uranium
intermetallic compound. In this context, we recall Zwicknagl and
Fulde's theory\cite{Zwicknagl} of the dual nature of $f$-electrons
in uranium compounds: they interpret the behaviour of $5f^3$ systems
such as UPt$_3$ in terms of two $5f$ electrons being localised and
one being itinerant.

Due to the inherent difficulties in handling neptunium, relatively
few neptunium compounds have been studied.  A principal motivation
for the work presented in this paper has been to explore the
magnetic and electrical properties of NpPd$_3$ with the aim of
comparing them with those of UPd$_3$, and to examine the effect of a
dilute substitution of U by Np on the unusual behaviour of UPd$_3$.

Numerous studies of UPd$_3$, both macroscopic (heat capacity
\cite{Zochowski95,Tokiwa}, susceptibility\cite{McEwen03}, thermal
expansion\cite{Zochowski94}, ultrasonics\cite{Lingg}) and
microscopic (neutron\cite{McEwen98} and X-ray
diffraction\cite{McMorrow}), indicate 4 phase transitions below
$T=8$ K. These have been attributed to a series of
antiferroquadrupolar (AFQ) orderings of the U $5f^2$ electrons.
However, a detailed understanding of the origin of these transitions
has been a challenge for many years, but a new proposed crystal
field scheme \cite{McEwen03} has been able to explain the existence
of the 4 transitions and new X-ray resonant scattering experiments
have revealed the order parameters of one of the AFQ phases
\cite{HWalkerPRL,McEwen07}. In comparison very little has been
published on the neptunium analogue DHCP $\mathrm{NpPd_3}$.  Its
magnetic properties were first examined 30 years ago by Nellis et
al. \cite{Nellis}, but several questions remain unanswered in
relation to the nature of the ordering. As far as the present
authors are aware, there has been no further research until our own
work, and we report here a systematic study of the magnetisation,
magnetic susceptibility, electrical resistivity and heat capacity of
NpPd$_3$. Given the very similar lattice parameters between the
isostructural U and Np compounds we have anticipated complete solid
solubility from UPd$_3$ to NpPd$_3$, and the very small difference
in ionic radii should lead to little induced strain. This has
enabled us to investigate the perturbation of the quadrupolar
structures of UPd$_3$ by substitution of Np on some U sites, whilst
looking to explain the nature of the ordering previously observed in
NpPd$_3$.

\section{Experimental Details}
\begin{figure}
\centering
        \includegraphics[width=0.45\textwidth,bb=15 15 310 210,clip]{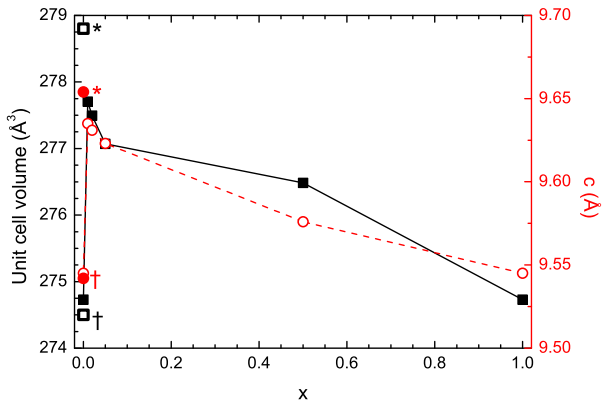}
        \caption{\label{vol}The variations of cell volume $(\blacksquare)$ and c-axis
        parameter $(\textcolor[rgb]{1.00,0.00,0.00}{\circ})$ as
        a function of $x$ in (U$_{1-x}$Np$_x$)Pd$_3$ do not obey Vegard's
        law. Data points taken from \cite{Kleykamp} for U rich ($*$)
        and Pd rich ($\dag$) UPd$_3$ samples are not included in the line guide.}
\end{figure}

Polycrystalline samples of DHCP (U$_{1-x}$Np$_x$)Pd$_3$ with $x=0$,
0.01, 0.02, 0.05, 0.50 and 1 were prepared at the Institute for
Transuranium Elements, by arc melting stoichiometric amounts of the
constituent elements in a high purity argon environment on a water
cooled copper hearth using a Zr getter.  We used Pd of $99.99\%$
purity, whilst our U and Np was of $99.9\%$ purity.  Quantitative
analysis of the impurities, including the oxygen content, was not
available to us. The resultant ingots were repeatedly turned and
remelted, before they were annealed at $1300^\circ$ C for one week
to ensure phase homogeneity. This is important as pure NpPd$_3$ is
known to exist in two phases: DHCP and cubic with the AuCu$_3$-type
structure \cite{Nellis}; the latter allotrope orders
antiferromagnetically at $T_N=50$ K. No weight losses were observed
during the arc melting and annealing processes, and given the low
vapour pressures of U, Np and Pd metals, the error bars on the
sample compositions are very small and correspond to the precision
of the balance used in the element weighing procedure, e.g. an error
of 3/1000 in weight and hence 4/1000 in atomic percent. The phase
purity of the samples was checked using room temperature X-ray
powder diffraction (Cu $K_\alpha$ radiation) on a Bragg-Brentano
D-500 diffractometer. A Rietveld-type full refinement of the data
proves that all the samples have the DHCP structure. The lattice
parameters of all the compounds are given in Table~\ref{tab:table1},
and the cell volume and c-axis parameter of (U$_{1-x}$Np$_x$)Pd$_3$
as a function of $x$ are plotted in Figure~\ref{vol}. Interestingly,
the cell volume rapidly increases with the addition of just $1\%$
neptunium before decreasing with further neptunium doping.  This is
not the standard behaviour and violates Vegard's law \cite{Vegard}.

\begin{table*}
\caption{\label{tab:table1}The structural parameters of DHCP
(U$_{1-x}$Np$_x$)Pd$_3$ compounds, where esd is the estimated
standard deviation.}
\begin{ruledtabular}
\begin{tabular}{cccccc}{Compound}
 &\multicolumn{3}{c}{Unit cell dimensions in {\AA} (esd$<0.001$)}&{Cell Volume in {\AA}$^3$} \\
  & a & c & c/a \\ \hline
UPd$_3$ & 5.765 & 9.545 & 1.656 & 274.7 & This work\\
 & 5.775 & 9.654 & 1.672 & 278.8 & U rich \cite{Kleykamp}\\
 & 5.763 & 9.542 & 1.656 & 274.5 & Pd rich \cite{Kleykamp}\\
 & 5.757 & 9.621 & 1.671 & 276.1 & Heal \& Williams\cite{Heal}\\
 & 5.769 & 9.652 & 1.673 & 278.2 & McEwen et al.\cite{McEwen88}\\
1\%Np & 5.769 & 9.635 & 1.670 & 277.7 \\
2\%Np & 5.768 & 9.631 & 1.670 & 277.5  \\
5\%Np & 5.766 & 9.623 & 1.669 & 277.1 \\
50\%Np & 5.774 & 9.576 & 1.658 & 276.4 \\
NpPd$_3$ & 5.765 & 9.545 & 1.656 & 274.7 & This work\\
 & 5.767 & 9.544 & 1.655 & 274.9 & Nellis et al.\cite{Nellis}
\end{tabular}
\end{ruledtabular}
\end{table*}

Also given in Table~\ref{tab:table1} are additional values for the
lattice parameters of UPd$_3$ from the literature, reflecting the
observation of a homogeneity range around the exact UPd$_3$
composition: $23.3$ to $24.8$ atomic percent uranium
\cite{Kleykamp}, which affects the lattice parameters significantly.
Based on these measurements it could appear that our UPd$_3$ sample
is slightly palladium rich. However, since the mixed actinide
samples were synthesised using the UPd$_3$ and NpPd$_3$ samples as
starting materials, in the correct stoichiometric ratios, any
uranium deficiency affecting the UPd$_3$ lattice parameters should
be preserved through the different composition samples. The sharp
deviation from a linear trend at low neptunium concentrations may be
possible evidence of a change in the valence state of uranium.

X-ray powder diffraction is unable, however, to show how the dopant
atoms sit within the host crystal structure.  We therefore do not
know whether the Np atoms in (U$_{1-x}$Np$_x$)Pd$_3$ are randomly
distributed over the two U sites (one locally hexagonal and the
other quasi-cubic) or whether they are located preferentially on one
of the symmetry type sites. An X-ray absorption spectroscopy
experiment is planned to look at the local configuration of the Np
atoms.

Measurements were made on encapsulated samples at ITU. Magnetic
studies were carried out for $T=2-300$ K and in fields up to 7 T
using a superconducting quantum interference device (SQUID)
magnetometer (Quantum Design MPMS-7). 4-point (pressure contacts) AC
probe electrical resistivity measurements were made using a Quantum
Design PPMS-9 for $T=2-300$ K in a range of fields up to 9 T.
Additional low temperature measurements were obtained using 2
coupled cryopump devices ($^3$He$-^4$He). Self heating effects make
it difficult to cool samples containing neptunium below 400 mK. Heat
capacity measurements were made also using the PPMS-9 via the hybrid
adiabatic relaxation method on both the (U$_{1-x}$Np$_x$)Pd$_3$
samples and a sample of ThPd$_3$, to be used as an isostructural
phonon blank. Using a $^3$He refrigeration insert, measurements
could be made over a range of $T=0.4-400$ K, in fields up to 9 T.
Contributions to the measured heat capacity from the sample platform
including the grease were measured separately and subtracted from
the total.  The heat capacity of the sample coating, stycast 2850
FT, is well known \cite{Javorsky} and was subtracted off from the
total heat capacity.

Hereafter the solid solutions investigated in this paper will be
identified, for ease and clarity, by the percentage neptunium
doping, for example 25\%Np would refer to
(U$_{0.75}$Np$_{0.25}$)Pd$_3$.

\section{Magnetic Properties}
\subsection{NpPd$_3$}
\begin{figure}
\centering
\subfigure{\label{npsus}\includegraphics[width=0.4\textwidth,bb=15
15 285 210,clip]{susnpv2.eps}}
\subfigure{\label{npinvsus}\includegraphics[width=0.4\textwidth,bb=15
15 285 210,clip]{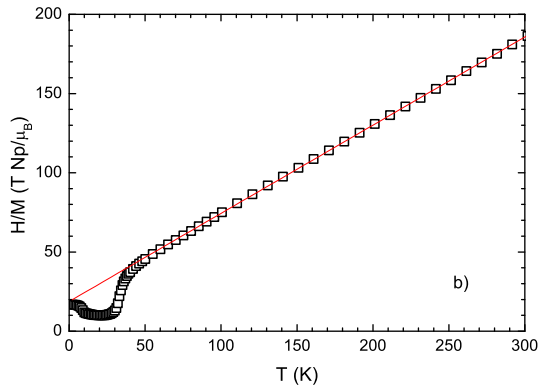}} \caption{\label{susc}(a) $M/H(T)$
of NpPd$_3$ at $H=1.1$ T ($\blacksquare$), 4 T ({\color{red}
$\circ$}) and 7 T ({\color{blue} $\blacktriangle$}), showing
transitions at $T = 10$ K and $30$ K. (b) Curie-Weiss fit to
$H/M(T)$ for $H=1.1$ T.}
\end{figure}

In NpPd$_3$, in low applied magnetic fields: 0.03 T and 1.1 T, $M/H$
(defined here for convenience as the magnetic susceptibility) shows
a sharp increase below 35 K to a broad maximum centred at 20 K, as
previously reported \cite{Nellis}. However, in addition we observe a
shoulder at 10 K. In higher fields of 4 T and 7 T, the form of $M/H$
changes such that the 10 K feature becomes more pronounced, see
Figure~\ref{susc}.

The 10 and 30 K anomalies could be attributed to two
antiferromagnetic transitions; these might occur separately on the
locally hexagonal and quasi-cubic sites, similar to the transitions
observed in neodymium \cite{Roberts}, or successively on only one of
the site types, as in praseodymium \cite{McEwen81}.  Whilst an increase of $M/H$
may be indicative of a ferromagnetic component, it is not inconsistent with the 
increases in $M/H$ seen at the successive antiferroquadrupolar 
transitions in UPd$_3$ \cite{McEwen07}. Interestingly,
no difference was seen in the neutron diffraction patterns measured
for dhcp NpPd$_3$ at $T=4.2$ K and $78$ K \cite{Nellis}, which could
be consistent with the higher temperature transition being to a
quadrupolar phase. This would result in a relatively high
quadrupolar temperature, considerably higher than those in UPd$_3$,
but comparable to that for DyB$_2$C$_2$: $T_Q=24.7$ K\cite{Yamauchi}
and for NpO$_2$: $T_Q=25$ K\cite{Paixao,comment}.  However, the
observed absence of magnetic Bragg peaks at 4.2 K suggested that
there is no long range magnetic order, which is inconsistent with
our proposed antiferromagnetic transition at 10 K.

The transition at $30$ K is seen more clearly in the inverse
susceptibility, see Fig.~\ref{npinvsus} . Above $50$ K the inverse
susceptibility follows a Curie-Weiss law, with an effective magnetic
moment of $2.83\pm0.05$ $\mu_B$ per Np atom deduced from a series of
measurements made in 1.1 T and 4 T. This suggests that Np is
trivalent in this compound, since in the Russell-Saunders coupling
scheme the moment value one would expect for a $5f^4$ electron
configuration is $2.68$ $\mu_B$, whereas if Np was tetravalent, as U
is in UPd$_3$, the expected moment would be $3.62$ $\mu_B$. Within
the intermediate coupling scheme, which given the high atomic mass
of the neptunium may be more appropriate, the expected moment for
trivalent Np is 2.755 $\mu_B$/Np ion, and tetravalent Np is 3.682
$\mu_B$/Np ion \cite{Fournier}, which again supports our conclusion
that Np is trivalent in NpPd$_3$.

\begin{figure}
\centering
\includegraphics[width=0.4\textwidth,bb=15 15 290 210,clip]{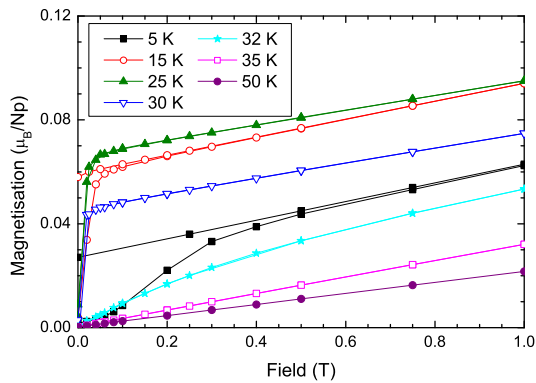}
\caption{\label{npmag}$M$ vs $H$ of NpPd$_3$ at a range of
temperatures showing hysteresis below 30 K.}
\end{figure}

The isothermal magnetization, Figure~\ref{npmag}, measured after
cooling the sample in zero field, below 30 K increases rapidly in
low fields before increasing more slowly and linearly.  On reducing
the field, hysteresis is observed below $0.1$ T, with a maximum
residual ferromagnetic moment of $0.06$ $\mu_B$/Np atom at $T=15$ K.
Below the 10 K transition, hysteresis is observed up to 0.5 T but
with a reduced residual ferromagnetic moment.  At 7 T the maximum
moment was $0.30$ $\mu_B$ per Np atom.  The absence of saturation in
7 T and the low remanent magnetisation, if intrinsic properties, may
indicate that the structure below 30 K contains a small
ferrimagnetic moment. An alternative explanation is the presence of
some ferromagnetic impurity phase, but the measurements made in
constant applied magnetic field appear to rule out either carbides
or nitrides, commonly occurring ferromagnetic impurities in other
systems, and the X-ray powder diffraction data shows the high purity
of the sample phase.  Nevertheless, a $2\%$ impurity of an
unidentified neptunium compound with a moment of 3 $\mu_B$/Np would
produce such a residual moment, and our X-ray powder diffraction
measurements cannot preclude such a possibility. However, we note
that no other Np-Pd binary compound has yet been reported.

Since neptunium is M\"{o}ssbauer active one might envisage measuring
the M\"{o}ssbauer spectra to learn more about the magnetic
properties of NpPd$_3$.  However, the data from previous
experiments\cite{Nellis,Gal} have proved difficult to interpret.
Although paramagnetic line broadening was observed below the 30 K
transition, at 4.2 K the data could not be fitted, even using two
hyperfine patterns taking into account the two inequivalent
neptunium sites.

\subsection{(U$_{0.5}$Np$_{0.5}$)Pd$_3$}
\begin{figure}
\centering
\subfigure{\label{sus50}\includegraphics[width=0.4\textwidth,bb=15
15 285 210,clip] {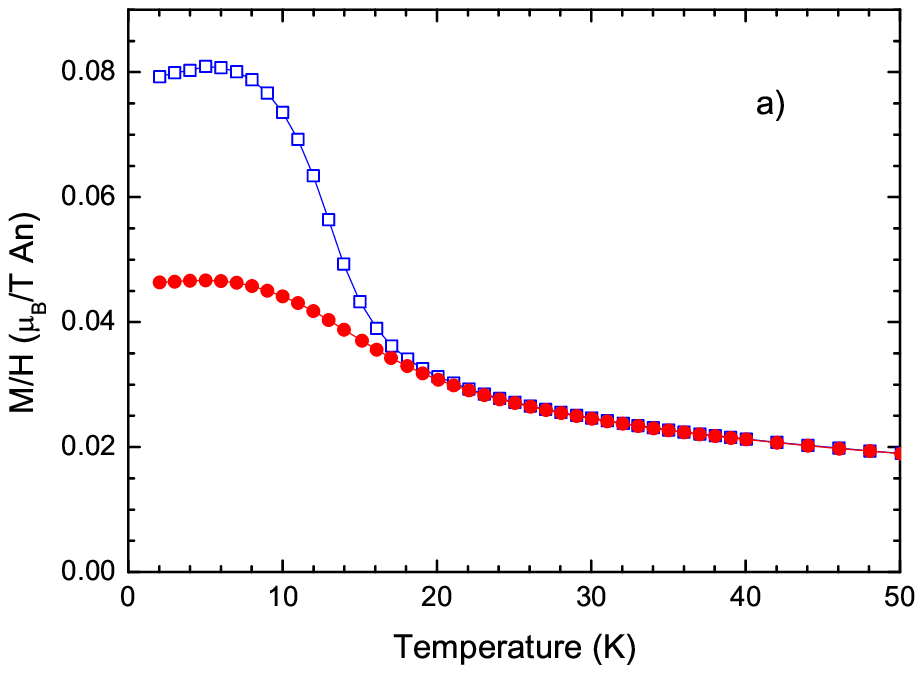}}\\
\subfigure{\label{isus50}\includegraphics[width=0.4\textwidth,bb=15 15 285 210,clip]
{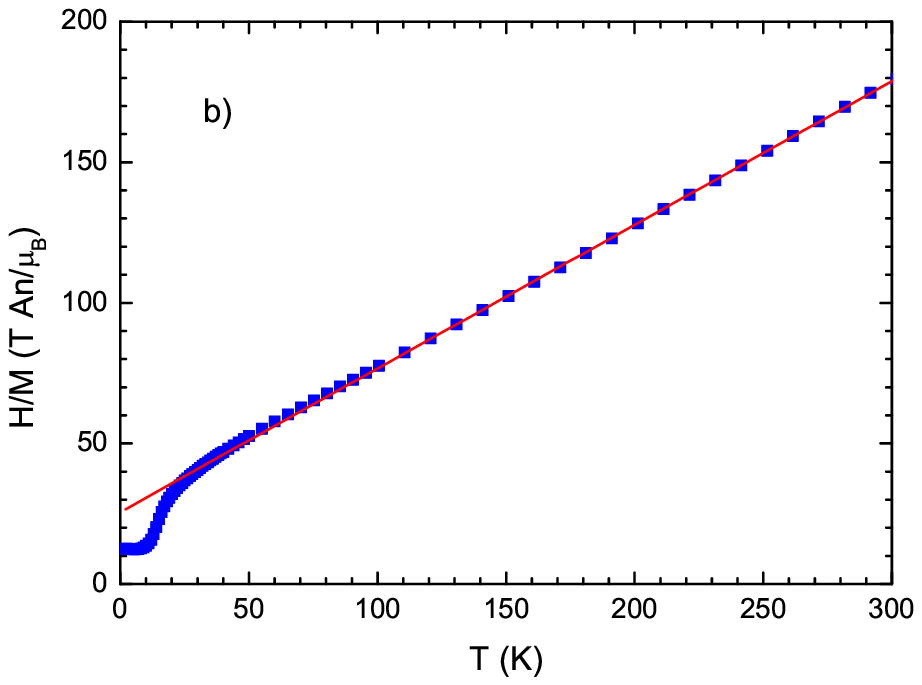}}
\caption{(a) $M/H(T)$ of
(U$_{0.5}$Np$_{0.5}$)Pd$_3$ at $H=1.1$ T ({\color{blue} $\square$})
and 4 T ({\color{red} $\bullet$}), showing a transition at $T \sim
12$ K. (b) Curie-Weiss fit to $H/M(T)$ for $H=1.1$ T giving an
effective moment of $2.956\pm0.003$ $\mu_B$/An.}
\end{figure}

The 50\%Np $M/H$ data, Figure~\ref{sus50} qualitatively resembles
that for pure NpPd$_3$ shifted down in temperature, such that there
is a sharp rise below 15 K to a broad maximum centred at $~5$ K.  In
the $H/M$ data the transition is seen clearly at 15 K, while above
the transition the data is linear exhibiting Curie-Weiss behaviour,
Figure~\ref{isus50}. Curie-Weiss fits to a series of data sets
measured in 1.1 T and 4 T give an effective paramagnetic moment of
$2.96\pm0.01$ $\mu_B$ per actinide ion. Hysteresis is observed in
magnetisation measurements as a function of applied magnetic field
while saturation is absent. Below 15 K the field below which
hysteresis is observed increases with decreasing temperature to 1 T
at 2 K.  The maximum residual ferromagnetic moment is $~0.05$
$\mu_B$/An.  Again we have not been able to distinguish whether it
is an intrinsic ferrimagnetic effect or if it is due to a very small
quantity of an impurity phase with a large ferromagnetic moment.
However, if it is assumed that the residual ferromagnetic moment
measured from the NpPd$_3$ sample was the result of a 2\%
unidentifed magnetic impurity phase, in this 50\%Np sample, which
was produced using the NpPd$_3$ sample, the impurity phase should
only make up 1\% of the total mass, and therefore one would have
expected the residual ferromagnetic moment to have halved, instead
of only being reduced by a sixth.

\subsection{Dilute Np samples, $x=0.01,\,0.02,\,0.05$}
\begin{figure}
\centering
\includegraphics[width=0.4\textwidth,bb=10 15 280 210,clip]{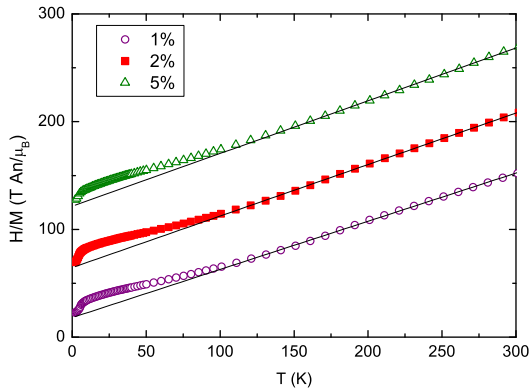}
\caption{\label{dilichi}$H/M$ measurements on dilute Np doped
samples for $H=1.1$ T with Curie-Weiss fits at high temperature. The
data sets have been displaced vertically by 50 and 100 units for
clarity.}
\end{figure}

The 5, 2 and 1\%Np magnetic measurements show no clear evidence of a
magnetic transition, and there is no evidence of hysteresis in the
isothermal magnetisation data.  Above 100 K, $H/M$ follows a
modified Curie-Weiss law for the 5\%Np sample, possibly indicative
of a singlet ground state with a large energy gap to the first
excited state, giving an effective paramagnetic moment of
$3.02\pm0.02$ $\mu_B$ per actinide ion. Fits to a Curie-Weiss law
for the 2\% and 1\%Np samples above 100 and 150 K give effective
paramagnetic moments of $3.07\pm0.01$ and $3.16\pm0.01$ $\mu_B$ per
actinide ion respectively, and are shown in Figure~\ref{dilichi}.

\begin{figure}
\centering
\includegraphics[width=0.4\textwidth,bb=15 15 280 210,clip]{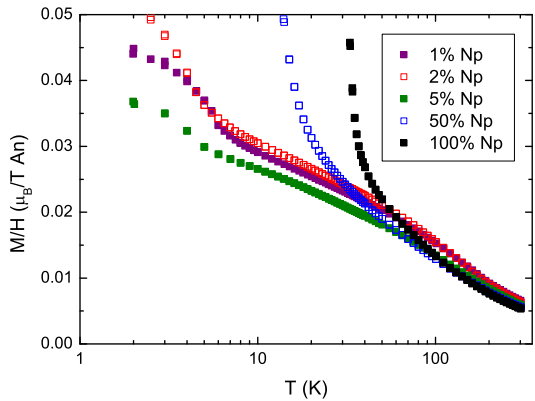}
\caption{\label{squidall}$M/H$ of pure NpPd$_3$ and mixed
(U,Np)Pd$_3$ at $H=1.1$ T plotted vs $\log T$.
(U$_{0.95}$Np$_{0.05}$)Pd$_3$ shows a near negative log trend,
possibly indicative of a non-Fermi Liquid.}
\end{figure}

Plotting the $M/H$ data for all the different samples: NpPd$_3$,
50\%Np, 5\%Np, 2\%Np and 1\%Np, as a function of temperature on a
logarithmic plot, Figure~\ref{squidall}, shows that the 5\%Np data
approximately follows a negative logarithmic trend. This can be
characteristic of non-Fermi Liquid behaviour. It is also interesting
to note that the value of $M/H$ at $T=2$ K is a minimum for
$x=0.05$, with the value increasing as a function of $x$ moving away
from this composition towards both the more dilute and more
concentrated regions of the phase diagram.

\begin{figure}
\centering
\includegraphics[width=0.4\textwidth,bb=15 15 280 230,clip]{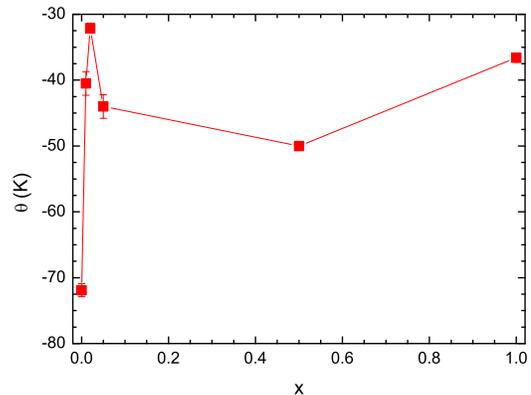}
\caption{\label{CWtemps}The Curie-Weiss temperature $\theta_P$ as a
function of $x$ in (U$_{1-x}$Np$_x$)Pd$_3$.}
\end{figure}

\begin{table*}
\caption{\label{tab:table2}Magnetic property parameters of DHCP
(U$_{1-x}$Np$_x$)Pd$_3$ compounds. Temperatures quoted in brackets
correspond to transitions which may be of quadrupolar origin.  The
effective magnetic moments quoted are the statistical averages from
Curie-Weiss fits to repeated inverse susceptibility data fits with
their associated errors.}
\begin{ruledtabular}
\begin{tabular}{cccc}
{Compound} & {Magnetic ordering temperatures
(K)}&{$\mu_\mathrm{eff}$ ($\mu_\mathrm{B}$/An ion)} &
{$\Theta_P$ (K)}\\
\hline
UPd$_3$ & - & $3.24\pm0.01$ & $-72\pm1$\\
1\%Np & - & $3.16\pm0.01$ & $-40\pm2$ \\
2\%Np & - & $3.06\pm0.01$ & $-32\pm1$ \\
5\%Np & - & $3.02\pm0.02$ & $-44\pm2$ \\
50\%Np & (12) & $2.96\pm0.01$ & $-50\pm1$ \\
NpPd$_3$ & 10 (30) & $2.83\pm0.05$ & $-35\pm1$ \\
\end{tabular}
\end{ruledtabular}
\end{table*}
Figure~\ref{CWtemps} shows how the Curie-Weiss temperature
$\theta_P$, obtained from Curie-Weiss fits to the high temperature
$H/M$ data and tabulated in Table~\ref{tab:table2}, varies as a
function of the neptunium concentration in (U$_{1-x}$Np$_x$)Pd$_3$.
The data point for $x=0$ was obtained from an appropriate average
($\frac{1}{3}\theta_P^{(c)}+\frac{2}{3}\theta_P^{(a)}$) of the
single crystal data \cite{McEwen03}. The sharp change in $\theta_P$
with low neptunium doping is very unusual. The behaviour may be
associated with the dramatic change in the lattice parameters,
Figure~\ref{vol}. One possibility is that the addition of a small
quantity of Np leads to a sharp change in the cohesive energy, and
hence the lattice constants, and the exchange interaction, which is
dependent on the conduction electrons, may be very sensitive to the
lattice spacing due to changing bands near the Fermi level.

\section{Electrical Resistivity}
\subsection{NpPd$_3$}
\begin{figure}
\centering \centering
\subfigure{\label{rho}\includegraphics[width=0.4\textwidth,bb=15 15
285 210,clip]{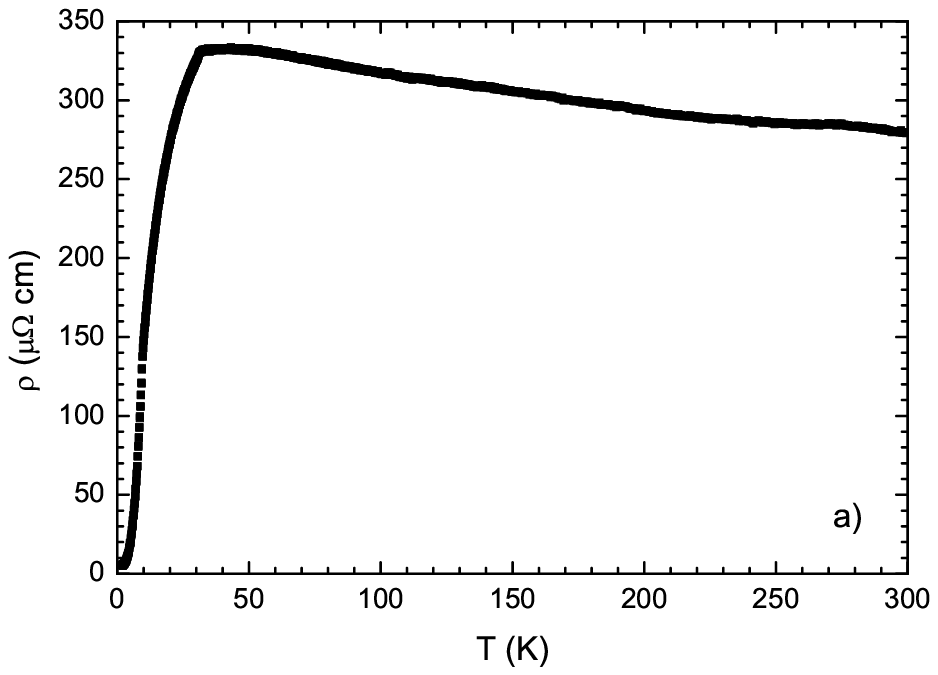}}
\subfigure{\label{rholowT}
\includegraphics[width=0.4\textwidth,bb=15
10 285 210,clip]{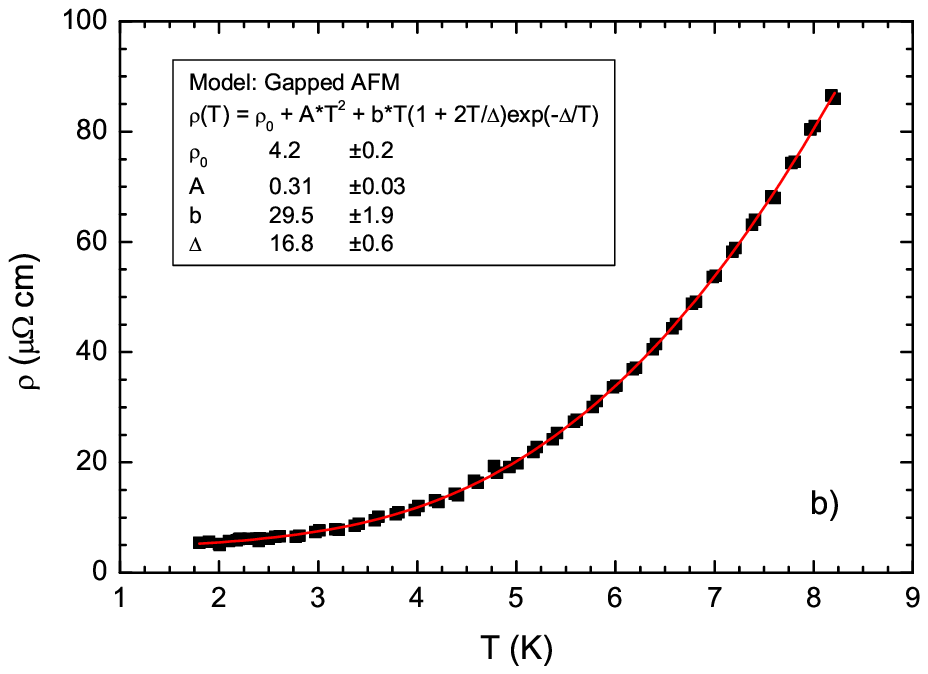}} \caption{(a) $\rho(T)$ of
NpPd$_3$ at $H=0$ T for $T=2-300$ K, showing a marked change in the
gradient at $T = 30$ K. Below 30 K the resistivity decreases with
decreasing temperature as for a normal metal, but above 30 K
$d\rho/dT$ is negative.  The 30 K transition appears to coincide
with the onset of coherence. (b) At low temperatures $\rho(T)$ is
not proportional to $T^2$. Instead it can be modelled using an
antiferromagnetic ground state with an energy gap $\Delta=17$ K
\cite{Andersen}.}
\end{figure}

In NpPd$_3$ in zero field the electrical resistivity shows two
anomalies, at $T=10$ K and 30 K, corresponding to the transitions
seen in the magnetisation data.  Below 300 K the resistivity
increases with decreasing temperature, see Figure~\ref{rho},
following a Kondo-like behaviour:
\begin{equation}
\rho=\rho_0+cT-\rho_S\ln(T),
\end{equation}
until 30 K, when it drops dramatically with the onset of coherent
scattering. In heavy fermion materials Kondo behaviour is commonly
observed in the form of a maximum in the resistivity at a
temperature $T_M$, which is a function of the Kondo temperature
$T_K$ and the mean RKKY interaction strength between the magnetic
ions\cite{Schilling}, rather than as a low temperature resistivity
minimum. For example, the resistivity of NpRu$_2$Si$_2$ is similar
to that of NpPd$_3$ displaying a logarithmic variation with T above
the ordering temperature, 27.5 K, and a precipitous drop below $T_M$
understood as a huge magnetic contribution of an energy gap
antiferromagnet\cite{Wastin}.  It is interesting that the higher
temperature transition observed in magnetisation measurements should
coincide with $T_M$.

The low temperature resistivity below 10 K does not vary as a simple
Fermi Liquid $\rho(T)\propto T^2$, but instead behaves as an
antiferromagnet with an energy gap $\Delta$:
\begin{equation}
\rho(T)=\rho_0+AT^2+bT(1+2T/\Delta)\mathrm{exp}(-\Delta/T)
\end{equation}
see Figure~\ref{rholowT}.  The fit to the $H=0$ T data indicates
$\Delta=17$ K. The residual resistivity ratio given by $\rho(T=0
K)/\rho(T=300 K)$ is $~85$, showing the high quality of the sample.

\begin{figure}
\centering
\includegraphics[width=0.4\textwidth,bb=15 15 285
210,clip]{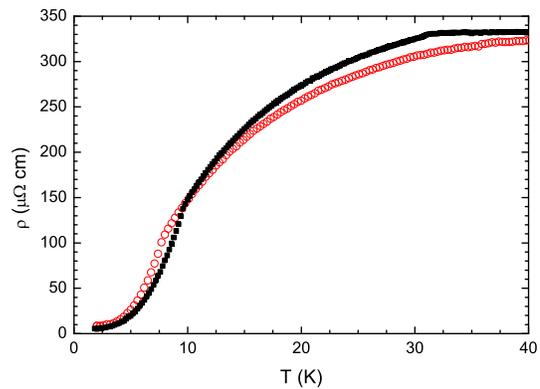}\caption{\label{rhofield} $\rho(T)$ for
NpPd$_3$ in $H=0$ T ($\blacksquare$) and 9 T ({\color{red}$\circ$}).
In zero field the two transitions at 10 and 30 K can be seen
clearly. In 9 T the upper transition is smoothed away, while the
lower transition is shifted down in temperature.}
\end{figure}

When a 9 T field is applied, the feature associated with the higher
temperature transition is smoothed away, while that of the lower
transition becomes more pronounced and shifts down in temperature,
consistent with an antiferromagnetic transition, see
Figure~\ref{rhofield}. Below the transition the 9 T data still
behaves as a gapped antiferromagnet but the parameters given by the
fit are modified to $A=0.14\pm0.04$ $\mu\Omega$cm/K$^2$ and
$\Delta=15.7\pm0.5$ K.

\begin{figure}
\begin{center}
\subfigure{\label{res50all}
\includegraphics[width=0.42\textwidth,bb=15 15 300
210,clip]{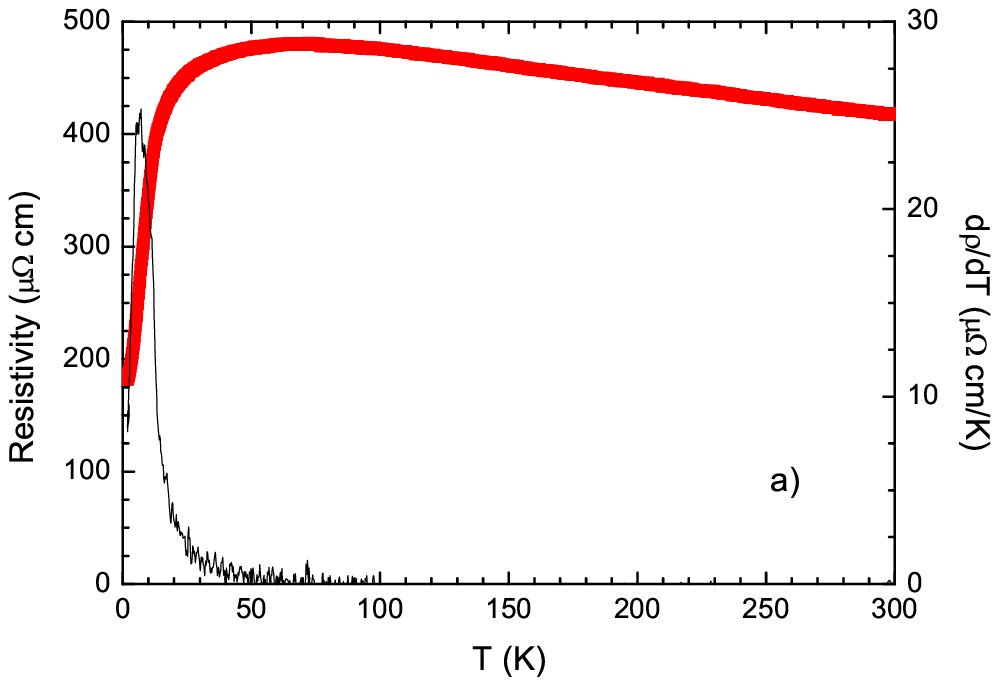}}\\
\subfigure{\label{res50low}\includegraphics[width=0.42\textwidth,bb=10
15 300 210,clip]{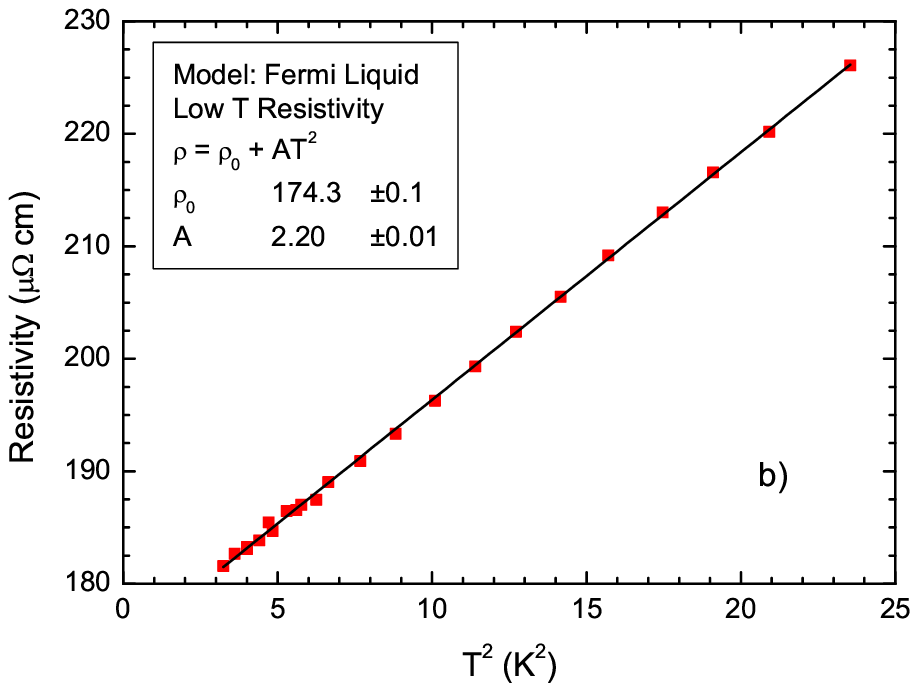}}
\end{center}
\caption{\label{res50} (a) $\rho(T)$ of (U$_{0.5}$Np$_{0.5}$)Pd$_3$
at $H=0$ T ({\color{red}$\blacksquare$}) for $T=2-300$ K, showing a
smooth change from a positive to a negative gradient at $T\sim50$ K.
The transition observed in the magnetisation measurements is seen
more clearly in the temperature derivative (black line). (b) Below 5
K the resistivity varies as $T^2$, in very good agreement with
Fermi-Liquid Theory.}
\end{figure}
\begin{figure}[h!]
\centering
\includegraphics[width=0.4\textwidth,bb=15 15 280
210,clip]{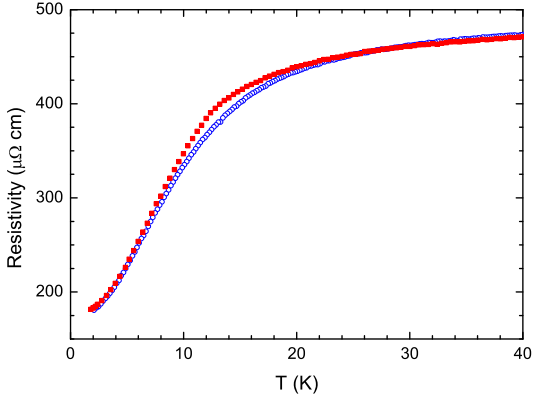} \caption{\label{res50field}$\rho(T)$ for
(U$_{0.5}$Np$_{0.5}$)Pd$_3$ in $H=0$ ({\color{red}$\blacksquare$})
and 9 ({\color{blue}$\circ$}) T. In zero field a transition can be
seen at 12 K. In 9 T the transition is smoothed away.}
\end{figure}
\subsection{(U$_{0.5}$Np$_{0.5}$)Pd$_3$}
Again, for the 50\%Np sample, the high temperature resistance has a
negative gradient, displaying a Kondo-like behaviour,
Figure~\ref{res50all}. However the onset of coherence is not so
sharp, varying over a broader range of temperatures, and is not
associated with a transition temperature deduced from the magnetic
measurements.  The low temperature resistivity varies as $T^2$, with
a residual resistivity of $174$ $\mathrm{\mu\Omega cm}$,
Figure~\ref{res50low}. This is some two orders of magnitude larger
than in pure NpPd$_3$, which we attribute to the statistical
disorder introduced with the uranium atoms.  As shown in
Figure~\ref{res50field}, a kink is observed in the zero field
resistivity at $T=12$ K, which is smoothed away in a 9 T field.

\begin{figure}
\centering
\subfigure{\label{rho5}\includegraphics[width=0.4\textwidth,bb=15 15
280 210,clip]{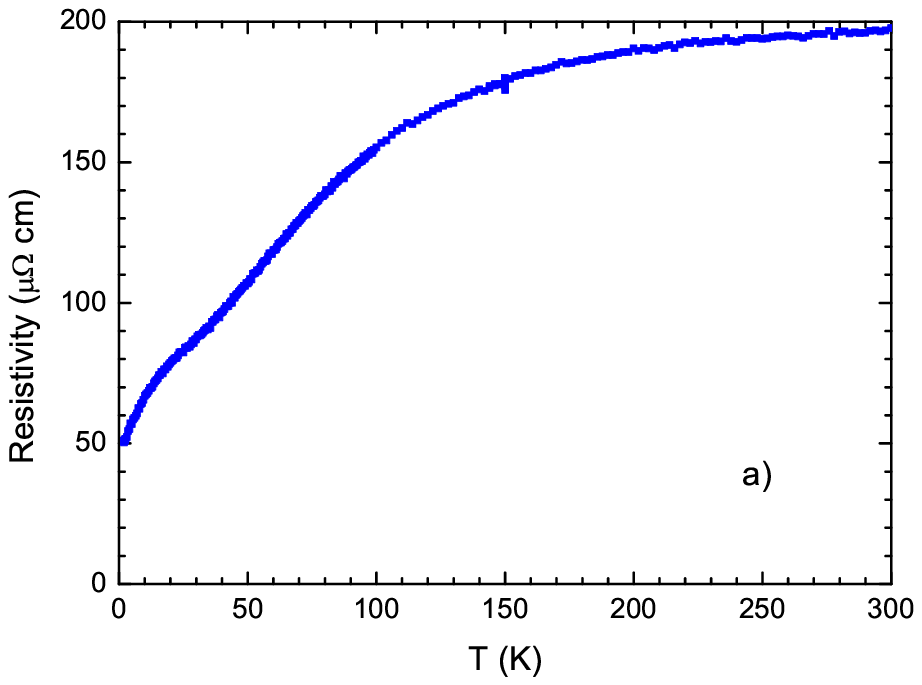}}
\subfigure{\label{res5low}\includegraphics[width=0.4\textwidth,bb=15
10 280 210,clip]{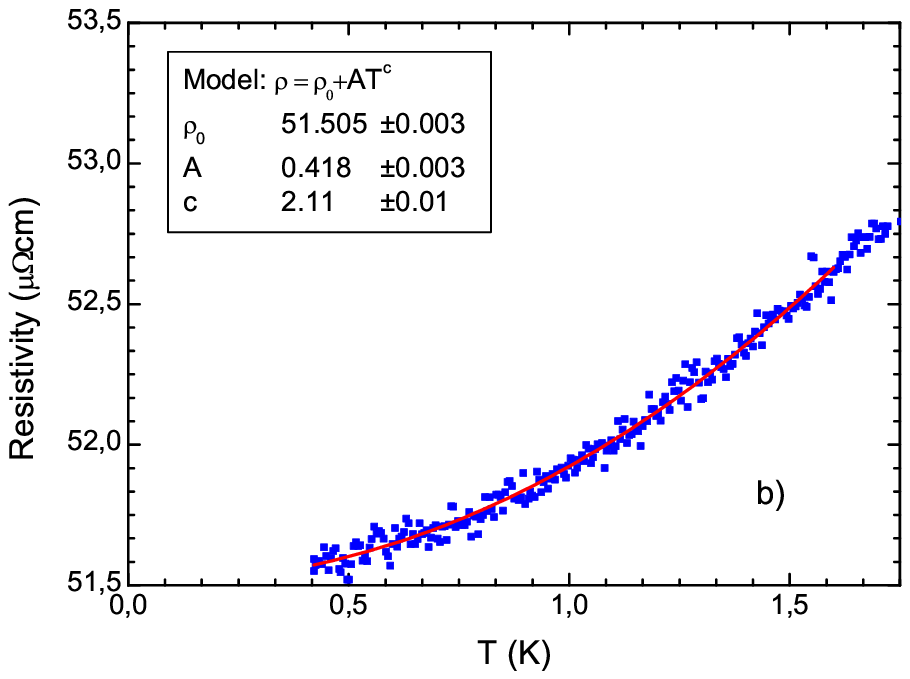}} \caption{(a)$\rho(T)$ of
(U$_{0.95}$Np$_{0.05}$)Pd$_3$ at $H=0$ T for $T=2-300$ K, showing a
standard metallic behaviour at high temperatures. (b) $\rho(T)$ at
$H=0$ T for $T=0.4-1.75$ K. A fit to $\rho=\rho_0+AT^c$ for
$T\leq1.6$ K indicates that this composition is a Fermi liquid.}
\end{figure}
\subsection{(U$_{0.95}$Np$_{0.05}$)Pd$_3$}
The electrical resistivity of the 5\%Np sample is markedly different
from the 50\% and 100\%Np samples showing no Kondo-like behaviour at
high temperatures, Figure~\ref{rho5}.  This raises the possibility
that the degree of electron localisation is varying as we move
across the phase diagram as a function of neptunium concentration.
Application of a 9 T field has little effect on the resistivity.
Initial resistivity measurements performed down to 2 K suggested a
linear temperature dependence for $2\leq T\leq10$ K. However, when
we extended these measurements to lower temperatures using the
coupled cryopump system, a fit to the data indicated that below 1.6
K the resistivity is quadratic in temperature, Figure~\ref{res5low},
indicating that at the lowest temperatures
(U$_{0.95}$Np$_{0.05}$)Pd$_3$ is a Fermi liquid.

\section{Heat Capacity}
\subsection{NpPd$_3$}
\begin{figure}
\centering
\includegraphics[width=0.4\textwidth,bb=15 15 285 210,clip]
{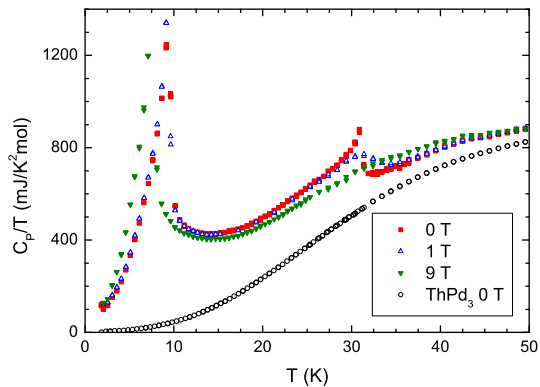} \caption{\label{nphc} $C_P/T$ vs $T$ of NpPd$_3$, showing
peak features at $T=10$ K and $T=30$ K in zero field, and their
evolution in different fields. Zero field results for ThPd$_3$ are
also shown.}
\end{figure}

In zero field, heat capacity measurements of NpPd$_3$ reveal two
clear lambda anomalies at $T=10$ and 30 K, Figure~\ref{nphc}. Making
a fit to the low temperature data gives the electronic heat
capacity, $\gamma=78\pm4$ mJK$^{-2}$mol$^{-1}$, and hence a
Kadowaki-Woods Ratio of $A/\gamma^2=5.2\pm0.8\times10^{-5}$
$\mu\Omega$cm(molK/mJ)$^2$, indicating that NpPd$_3$ is possibly a
moderately heavy Fermion material. However, estimating $\gamma$ is
complicated by the significant curvature due to the lambda feature
at 10 K, such that this value for the electronic heat capacity
corresponds to a fit: $C=\gamma T + \beta T^3$ which gives a value
for $\beta$ such that the Debye temperature would be 61 K, which is
clearly too low.

In increasing applied magnetic fields the anomaly associated with
the 10 K transition shifts down in temperature, in agreement with
the behaviour expected from an antiferromagnetic transition. The
feature also decreases in size until $H=7$ T after which it sharpens
and increases in magnitude, behaviour which is as yet not explained.
The application of a 1 T field broadens the higher temperature
transition feature and  shifts it up in temperature by $~2$ K, while
in fields greater than 4 T the feature has been smoothed away. These
are not the characteristics of an antiferromagnetic phase
transition, however this could be consistent with a quadrupolar
transition.

We have measured the heat capacity of ThPd$_3$ under identical
experimental conditions.  Using this isostructural non-magnetic
compound as a phonon blank, the magnetic entropy of NpPd$_3$ has
been calculated. It reaches $R\ln4$ at 40 K, which suggests low
lying excited doublet crystal field states above ground state
doublets on both the locally hexagonal and quasi-cubic sites.

\subsection{(U$_{0.5}$Np$_{0.5}$)Pd$_3$}
\begin{figure}
\centering
\includegraphics[width=0.4\textwidth,bb=15 15 290 210,clip]{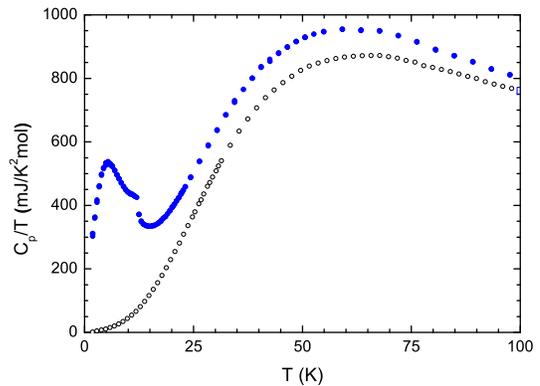}
\caption{\label{np50cp}$C_P/T$ for (U$_{0.5}$Np$_{0.5}$)Pd$_3$
({\color{blue}$\bullet$}) and ThPd$_3$ ($\square$) at $H=0$ T for
$T=2-100$ K.}
\end{figure}

The heat capacity data for 50\%Np shows no lambda-like anomalies.
Instead, inspection of Figure~\ref{np50cp} reveals a shoulder kink
at 12 K in zero field in $C_P/T$, corresponding to the temperature
at which features are observed in the magnetisation and resistivity
measurements, and a broad peak centred on 5 K.  No features are seen
at 5 K in the susceptibility or resistivity measurements in this
composition. This may be the Schottky peak arising from the
splitting of the ground state doublet in the molecular field below
the 12 K ordering temperature. Application of a 2 T field smoothes
away the 12 K feature, while fields up to 9 T make the 5 K peak
slightly narrower, but it does not change in magnitude.

The curvature of $C_P/T$ down to 2 K is such that it is very
difficult to make an estimate for the electronic heat capacity.  It
is quite possible that there is another transition below 2 K, but
neglecting this possibility and extrapolating the data down to zero
temperature suggests that $\gamma$ is in the region of $150-250$
mJK$^{-2}$mol$^{-1}$, considerably larger than that for NpPd$_3$ or
UPd$_3$. This results in a value of
$A/\gamma^2=5.5\pm2.8\times10^{-5}$ $\mu\Omega$cm(molK/mJ)$^2$,
which may indicate that 50\% Np is another heavy fermion material. A
possible explanation is that the Np doping induces a valence change
on the U ions leading to a lowering of the Fermi Energy closer to
the \emph{f} bands, resulting in a greater deal of hybridization and
hence heavy fermion behaviour. Without an accurate estimate for
$\gamma$ it is not possible to calculate accurately the magnetic
entropy of the sample.

\subsection{(U$_{0.95}$Np$_{0.05}$)Pd$_3$}
\begin{figure}
\centering
\includegraphics[width=0.4\textwidth,bb=15 15 280
210,clip]{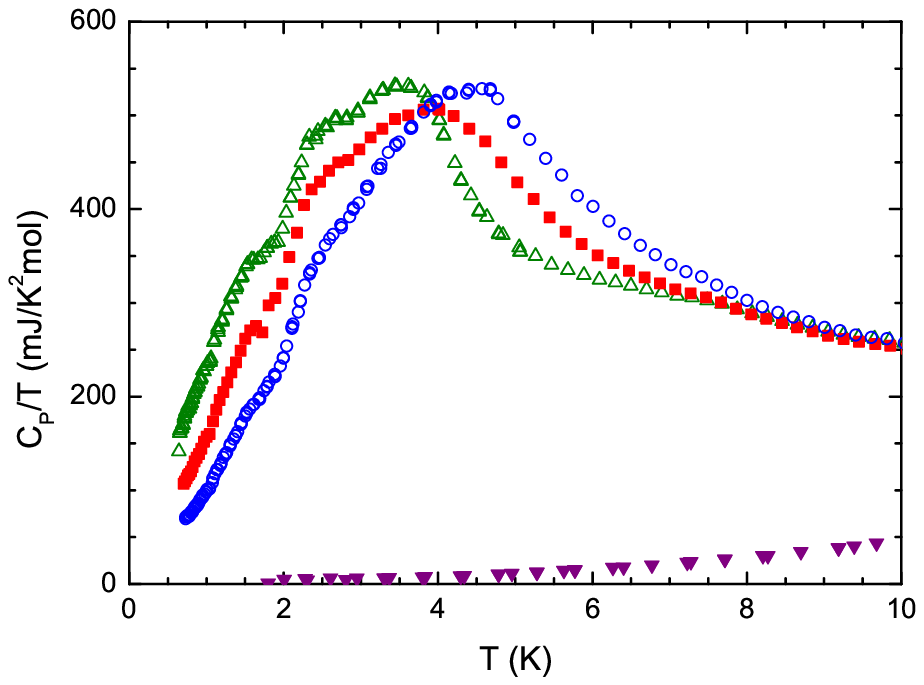} \caption{\label{HC5} $C_P/T$ in
(U$_{0.95}$Np$_{0.05}$)Pd$_3$ for $H=0$ T
({\color[rgb]{0.00,0.40,0.29}$\vartriangle$}), $H=5$ T
({\color{red}$\blacksquare$}) and $H=9$ T, ({\color{blue}$\circ$})
and in ThPd$_3$
({\color[rgb]{0.5,0.00,0.5}{$\blacktriangledown$}}).}
\end{figure}
Initial $C_P/T$ measurements of 5\%Np in zero field showed a broad
rounded peak centred on 3.5 K with a value of $400$
mJK$^{-2}$mol$^{-1}$ at 2 K. The significant curvature at this
temperature made extrapolating $C_P/T$ back to $T=0$ K very
difficult. So to determine whether there are additional transitions
which lead to a reduced estimate for $\gamma$, or whether this is
another heavy fermion, as might be associated with proximity to a
quantum critical point due to the suppression of a quadrupolar
transition to zero temperature, additional measurements were
performed using a $^3$He insert. These revealed the existence of
more features in the data at 2.5 K and 1.5 K, which may be
associated with further transitions, see Figure~\ref{HC5}. The
presence of the transitions at such low temperatures makes
estimating $\gamma$ very difficult, but assuming there are no
further transitions and continuing the trend in $C_P/T$ gives an
estimate of $\gamma=125\pm25$ mJ/K$^2$mol. This results in a
Kadowaki-Woods ratio of $2.7\pm1.1\times10^{-5}$
$\mu\Omega$cm.(mol.K/mJ)$^2$, indicating that the compound may also
be a heavy fermion. Application of increasing magnetic fields
results in the features shifting up in temperature. Such behaviour
is either indicative of ferromagnetic transitions, in which field
reinforces the magnetic structure, or quadrupolar transitions. Since
no hysteresis is seen in the magnetisation down to 2 K,
ferromagnetism has been excluded as a possible explanation.

\subsection{(U$_{1-x}$Np$_x$)Pd$_3$, $x=0.01,\,0.02$}
\begin{figure}
\centering
\includegraphics[width=0.4\textwidth,bb=15 15 285 210,clip]{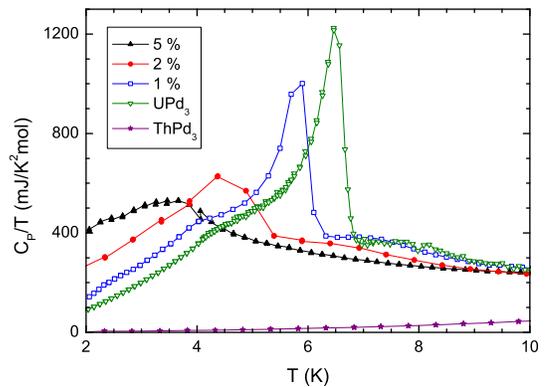}
\caption{\label{cpdil}$C_P/T$ of (U$_{1-x}$Np$_x$)Pd$_3$, with
$x=0,\,0.01,\,0.02,\,0.05$, and ThPd$_3$ in zero applied magnetic
field, showing the suppression of the quadrupolar transition
temperatures with increasing Np concentration.}
\end{figure}

\begin{figure}
\centering
\includegraphics[width=0.4\textwidth,bb=18 15 285
210,clip]{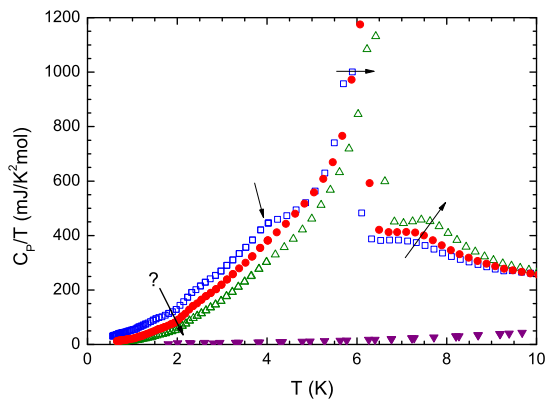}\caption{\label{HC1} $C_P/T$ of
(U$_{0.99}$Np$_{0.01}$)Pd$_3$ for $H=0$ T ({\color{blue}$\square$}),
$H=5$ T ({\color{red}$\bullet$}) and $H=9$ T
({\color[rgb]{0.00,0.40,0.29}$\vartriangle$}), and in ThPd$_3$
({\color[rgb]{0.5,0.00,0.5}{$\blacktriangledown$}}). The general
shape is very reminiscent of that for pure UPd$_3$.  The arrows show
how the features associated with transitions evolve as a function of
the applied field. The feature labeled with a question marked arrow
may not be due to an intrinsic property of the system.}
\end{figure}

As one can see in Figure~\ref{cpdil}, the heat capacity of 2\%Np and
1\%Np shows strong similarities to the data for UPd$_3$
\cite{HWalkerPRL}.  This observation is difficult to reconcile with 
the hypothesis that the uranium valence is changing with low neptunium 
doping, as postulated from the anomalous behaviour of the lattice 
parameters as shown in Figure~\ref{vol} and discussed in Section II.
It is also clear that the
heat capacity of ThPd$_3$ below 10 K is much less than in the $5f$
analogues. The 2\%Np $C_P/T$ data have a peak at $T=4.2$ K and a
shoulder at 7 K, features qualitatively similar to those at 6.5 and
7.8 K in polycrystalline UPd$_3$, though the peak is approximately
half the size in 2\%Np. When a field is applied the features
relating to the transitions shift up in temperature, such that in
$H=9$ T the peak is at 5.5 K. In addition the shoulder becomes more
pronounced with increasing field. The 1\%Np $C_P/T$ data, including
additional measurements made using a $^3$He insert, is shown in more
detail in Figure~\ref{HC1}, revealing transitions at 4, 6 and 7.3 K,
which may be labelled as $T_0$, $T_{-1}$ and $T_2$ using the
nomenclature for UPd$_3$ \cite{HWalkerPRL}. There is another feature
in the data at 2 K, which remains almost unchanged with increasing
applied magnetic field. The origin of this feature is as yet
unresolved. As for the 5\%Np sample, the positive shift of features
in temperature with field for 1\% and 2\%Np samples without
hysteresis observed in magnetisation measurements, in combination
with their qualitative resemblance to those in UPd$_3$, leads us to
believe that these are transitions to quadrupolar phases.

\section{Conclusions}
In conclusion, magnetisation, electrical resistivity and heat
capacity measurements reveal two transitions in dhcp NpPd$_3$ at
$T_2=10$ K and $T_1=30$ K.  The in-field behaviour of the features
associated with the $T_2$ transition in different measurements
indicates that it is most probably an antiferromagnetic transition.
The $T_1$ transition is that which was observed in the previous
experiments by Nellis \emph{et al}., \cite{Nellis} but unattributed.
It has proved more difficult using the current polycrystalline bulk
property experimental data to assign the nature of this transition.
Initially, based on the magnetisation data, $T_1$ was also assessed
as being an antiferromagnetic transition \cite{HWalker}; however,
the in-field behaviour seen in resistivity and heat capacity results
is inconsistent with such a conclusion.  The smoothing away of the
features with increasing field is more reminiscent of a quadrupolar
transition. Interestingly the $T_1$ transition is at $T_M$ in the
resistivity data, which shows high temperature Kondo behaviour.

We plan to use neutron scattering to try to determine the nature of
the two transitions. The previous diffraction experiment
\cite{Nellis} observed no additional peaks at 4.2 K. However, the
experiment was performed on a very small quantity of polycrystalline
NpPd$_3$ with poor instrumental flux and resolution. With the
immense improvements in neutron sources and instrumentation that
have taken place over the past thirty years we are confident that
new diffraction data below 10 K will provide information about the
ordering vector of the antiferromagnetic phase.  Measurements in the
phase between the two transitions at $T=10$ K and 30 K will be
particularly interesting, since the absence of magnetic Bragg peaks
would point to a non-magnetic origin for the transition at 30 K.
Neutrons do not couple directly to quadrupole moments, but instead
we may observe a lattice modulation vector associated with an
induced lattice distortion as is seen in UPd$_3$. Polarised neutron
diffraction techniques may be used to follow the individual
susceptibilities of the hexagonal and quasi-cubic sites to
distinguish between magnetic moments ordering on the different site
types successively or simultaneously.

Our magnetic susceptibility measurements have demonstrated that
neptunium is trivalent in NpPd$_3$ in contrast to uranium which is
tetravalent in UPd$_3$, indicating a valence transition moving
across the phase diagram. This underlines the unusual properties of
UPd$_3$ since uranium is most commonly trivalent and raises the
possibility of a tendency towards a valence instability, such that a
critical percentage neptunium doping could induce a valence change on
the uranium ion.

\begin{figure}
\centering
\includegraphics[width=0.4\textwidth,bb=125 285 455 540,clip]{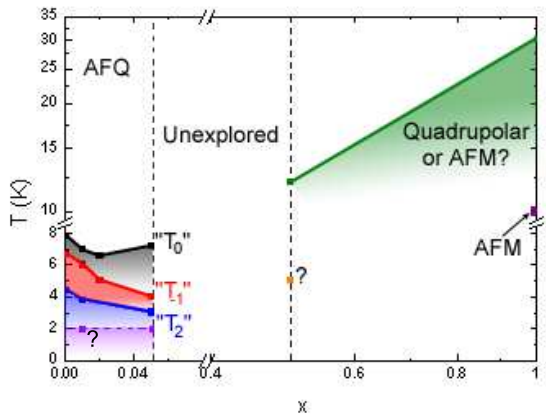}
\caption{\label{UNpphases}Temperature $T$ versus Neptunium
concentration $x$ phase diagram for (U$_{1-x}$Np$_x$)Pd$_3$ obtained
from bulk property measurements. Points marked with ``?'' are from
features in the heat capacity data, which either have yet to be
assigned to a transition, or are considered dubious. AFQ:
antiferroquadrupolar, AFM: antiferromagnet.}
\end{figure}

The results relating to all the different measurements performed on
polycrystalline samples of (U$_{1-x}$Np$_x$)Pd$_3$ can be summarised
in the phase diagram in Figure~\ref{UNpphases}. In the dilute Np
region of the phase diagram quadrupolar order is observed. The
dominant feature in the heat capacity of UPd$_3$ is the lambda
anomaly at $T_{-1}$, and so the evolution of this feature is the
easiest to follow with increasing Np doping. The ``$T_{-1}$''
feature moves to lower temperatures as $x$ is increased to $0.05$,
suggesting the possibility that at some critical doping
concentration this transition may be suppressed to zero temperature
leading to a quantum critical point of quadrupolar origin.

Proximity to a quantum critical point at $x=0.05$ could be
consistent with the apparent heavy fermion and possible non-Fermi
liquid behaviours at this composition observed in the bulk
thermodynamic measurements.  It would be very interesting to
discover a QCP due to the suppression of quadrupolar order, but one
has to query the nature of any critical fluctuations with which it
would be associated. The possibility of $E/T$ scaling in samples
close to criticality may be investigated using inelastic neutron
scattering.


Further investigation of the location of the dopant Np ions in
dilute (U$_{1-x}$Np$_x$)Pd$_3$ is desirable to determine wheter or
not they are distributed randomly over the two sites of the dhcp
structure.

The results presented in this paper provide an extensive survey of
the bulk magnetic and electrical properties of dhcp NpPd$_3$ and a
preliminary characterisation of the new and interesting mixed
actinide system (U,Np)Pd$_3$. Clearly there is considerable scope
for further experimental investigation.

\begin{acknowledgments}
H.C.W. thanks EPSRC for a research studentship and the Actinide User
Lab at ITU. We acknowledge the financial support to users provided
by the European Commission, DG-JRC within its ``Actinide User
Laboratory'' programme, and the European Community-Access to
Research Infrastructures action of the Improving Human Potential
Programme (IHP), contract HPRI-CT-2001-00118, and contract
RITA-CT-2006-026176. We would like to acknowledge helpful
discussions with G.H. Lander. The high purity Np metal used in this
work was made available through a loan agreement between Lawrence
Livermore National Laboratory and ITU, in the frame of a
collaboration involving LLNL, Los Alamos National Laboratory and the
US Department of Energy.
\end{acknowledgments}

\bibliography{paper}

\end{document}